\documentstyle[12pt]{article} 
\begin{document}     
\def\o{\over}
\def\Ar{\rightarrow}
\def\bar{\overline}
\def\r{\gamma}
\def\d{\delta}
\def\a{\alpha}
\def\b{\beta}
\def\n{\nu}
\def\m{\mu}
\def\k{\kappa}
\def\e{\epsilon}
\def\p{\pi}
\def\th{\theta}
\def\om{\omega}
\def\vp{{\varphi}}
\def\Re{{\rm Re}}
\def\Im{{\rm Im}}
\def\ra{\rightarrow}
\def\t{\tilde}
\def\bar{\overline}
\def\l{\lambda}
\def\G{{\rm GeV}}
\def\M{{\rm MeV}}
\def\eV{{\rm eV}}
%
\setcounter{page}{1}
\thispagestyle{empty}
\begin{flushright}    
{EHU-99-10, October 1999}\\
{hep-ph/9910261}\\
\end{flushright}   
\vskip 1 cm
\centerline{\LARGE \bf Neutrino Masses and Mixings}
\vskip 0.5 cm
\centerline{\LARGE\bf  with Flavor Symmetries
\footnote{To appear in the proceedings of  
"the XXIII International School of Theoretical Physics 
(Recent developments in theory of fundamental interactions)" at Ustron (Poland).}}
\vskip 2 cm
\centerline{{\large \bf Morimitsu TANIMOTO}
  \footnote{E-mail address: tanimoto@edserv.ed.ehime-u.ac.jp} }
\vskip 1 cm
 \centerline{ \it{Science Education Laboratory, Ehime University, 790-8577 Matsuyama, JAPAN}}
\vskip 3 cm
\centerline{\large \bf Abstract}
\vskip 1 cm 
Recent atmospheric neutrino data at Super-Kamiokande suggest the large
 flavor mixing of neutrinos.
  Models for the lepton  mass matrix, which give the near-maximal flavor mixing,
  are discussed in the three family model. Especially, 
     details of the models with the  $S_3$ or $O(3)$ flavor symmetry are studied.
\newpage
\section{Introduction}

Recent experimental data of neutrinos make big impact on the neutrino masses and their mixings.
Most exciting one is the results at Super-Kamiokande on the atmospheric neutrinos,
 which indicate the large neutrino flavor oscillation of $\n_\m\Ar \n_\tau$ \cite{SKam}.
 Solar neutrino data also provide the evidence of the neutrino oscillation, however,
 this problem is still uncertain \cite{BKS}. 
 
 Furthermore, a new stage is represented by the long baseline(LBL) neutrino oscillation
  experiments.
   The first LBL  reactor experiment CHOOZ has 
  provided a  bound  of the neutrino oscillation \cite{CHOOZ},
    which gives a strong constraint of the flavor mixing pattern.
 The LBL accelerator  experiment K2K \cite{K2K}  begins taking data,  
 whereas the MINOS \cite{MINOS} and ICARUS \cite{ICARUS}
  experiments will start in the first year of the next century. Those LBL experiments are 
  expected to clarify  masses, flavor mixings and $CP$ violation of neutrinos. 

The short baseline experiments may be helpful to understand  neutrino masses and
 flavor mixings.
	The tentative indication has been already given by the LSND experiment \cite{LSND},
   which is an accelerator experiment for $\n_\m\Ar\n_e(\bar \n_\m\Ar \bar \n_e)$.
   The  CHORUS  and NOMAD experiments \cite{CHO,NOM}
   have  reported the  new bound for $\n_\m\Ar\n_\tau$ oscillation,
   which has already improved the E531 result \cite{E531}.
   The  KARMEN experiment \cite{KARMEN}
 is also searching for the $\n_\m\Ar\n_e(\bar \n_\m\Ar \bar \n_e)$ oscillation 
 as well as LSND. However, they did not observed any evidences of the oscillation.
 The Bugey \cite{Bugey} and Krasnoyarsk \cite{Kras} reactor experiments and 
	  CDHS \cite{CDHS} and CCFR \cite{CCFR} accelerator experiments have given
	bounds for the neutrino mixing parameters as well as E776 \cite{E776}.
  
 What can we learn from these experimental results?
 We want to get clues for  the origin of neutrino masses and neutrino flavor mixings.
 In this paper, we concentrate our discussion on the flavor symmetry, 
 which controls the flavor structure of quark-lepton masses and mixings.

 \section{Possible Neutrino Mass Hierarchy and Mass Matrix Texture} 
 
 Our starting point as to the neutrino mixing is
  the large $\nu_\m \Ar \nu_\tau$ oscillation of the atmospheric neutrino oscillation with 
 $\Delta m^2_{\rm atm}=  (2\sim 6)\times  10^{-3} \eV^2$ and 
 $\sin^2 2\th_{\rm atm} \geq 0.84$,
 which are derived from the recent data of the atmospheric neutrino deficit at Super-Kamiokande
   \cite{SKam}. 
   In the solar neutrino problem \cite{BKS},
 there are three solutions:
  the small or large  mixing angle MSW \cite{MSW} solution  and 
  the vacuum oscillation solution (just so solution) \cite{BPW}.
  These mass difference scales are much smaller than the atmospheric one.
  
  Once we put $\Delta m_{\rm atm}^2=\Delta m_{32}^2$ and $\Delta m_{\odot}^2=\Delta m_{21}^2$, 
 there are three typical mass patterns: $m_3\gg m_2 \geq m_1$,
 $m_3\simeq m_2\simeq m_1$ and $m_1\simeq m_2 \gg m_3$.
 In this case, the LSND data is disregarded because there are only two mass difference scales
 in the three family model.

  If one goes to beyond three neutrinos, the sterile neutrinos are introduced.  These
 reconcil LSND result \cite{LSND}. 
   Then one can explain the difference of the mixing pattern  between quarks and leptons
   because  the sterile neutrino couples to  active neutrinos.
   This case has been discussed by Grimus in this school \cite{Grimus}.
 
 The neutrino mixing is  defined  as
  $\n_\a=U_{\a i}  \n_i$ \cite{MNS},      
   where $\a$ denotes the flavor $e,\m,\tau$  and  $i$ denotes mass eigenvalues $1,2,3$.
  Now we have two typical mixing patterns:
  \begin{equation}
   U_{\rm MNS} \simeq \left (\matrix{ 1 & U_{e2} & U_{e3} \cr
                          U_{\m 1} & {1\o \sqrt{2}} & -{1\o \sqrt{2}} \cr
		 U_{\tau 1} & {1\o \sqrt{2}} & {1\o \sqrt{2}} \cr	} \right )     ,
		 \qquad \qquad
	\left (\matrix{ {1\o \sqrt{2}} & -{1\o \sqrt{2}} & U_{e3} \cr
                         {1\o 2} & {1\o 2} & -{1\o \sqrt{2}} \cr
		  {1\o 2} & {1\o 2} & {1\o \sqrt{2}} \cr	} \right )     ,
 \end{equation}
  \noindent  the first one is the single maximal mixing pattern, in which
   the solar neutrino deficit is explained by the small mixing angle
   MSW solution,  and the other is the bi-maximal mixings pattern \cite{Bimax}, in which 
  the solar neutrino deficit is explained
   by the just so solution or the large mixing angle of MSW solution. 
    In both case $U_{e3}$ is constrained by the CHOOZ data \cite{CHOOZ}.
	

Before discussing possible mass matrices of neutrinos, we show how to
get $U_{\rm MNS}$ from the mass matrix as follows:
\begin{equation}
   U_{\rm MNS} = L_E^\dagger L_\nu \ ,  \quad m_E^{diagonal}=L_E^\dagger m_E R_E \ ,
                                    \quad m_\nu^{diagonal}=L_E^T m_\nu L_E \ ,
\end{equation}
\noindent where neutrinos are assumed to be Majorana particles.
So the large mixing in $U_{\rm MNS}$ could come from $L_E^\dagger$ or/and  $L_\nu$.
The pattern of the  $2\times 2$ sub-matrix  with $\sin^2 2\theta = 1$ are given 
in terms of the small parameter $\e$ as
\begin{eqnarray}
 &&\left (\matrix{ 1 & 1 \cr 1 & 1 \cr } \right )   \Longrightarrow  ( 0, 2) \ , \qquad
 \left (\matrix{ 1 & \e \cr \e & 1 \cr } \right ) \Longrightarrow  ( 1-\e, 1+\e) \ , \nonumber\\
 &&\left (\matrix{ \e & 1 \cr 1 & \e \cr } \right ) \Longrightarrow  ( -1-\e, 1-\e) \ , \qquad 
 \left (\matrix{ \e & 1 \cr \e & 1 \cr } \right ) \Longrightarrow  ( 0, \sqrt{2(1+\e^2)}) \ . 
\end{eqnarray}
 The first matrix gives the hierarchical eigenvalues, so it is useful for
 the neutrino and charged lepton mass matrices.
 The second and third ones give almost degenerated eigenvalues, which are
 useful only for neutrino masses.
 The last one is the aymmetric mass matrix with the hierarchical eigenvalues.
 So it is  useful only for the charged lepton.
 If the $3\times 3$ mass matrix includes these sub-matrices, the maximal mixing
 is derived. Moreover, there are some additional patterns in the $3\times 3$ matrix \cite{Alt}.

 The left handed neutrino masses are supposed to be at most ${\cal O}(1) \eV$.
We need some physical reasons  for the smallness of the neutrino mass.
In the case of Majorana neutrino, we know two classes of models 
which lead naturally to a small neutrino mass:
(i) models in which the seesaw mechanism works \cite{seesaw}
 and (ii) those in which the neutrino mass is  induced by a radiative correction.  
The central idea of  models (i) supposes some higher symmetry 
which is  broken at an high energy scale. If this symmetry breaking takes place 
so that it allowes the right-handed  neutrino to have a mass, and a small mass induced
for the left handed neutrino by the seesaw mechanism. 
 In the classes of model (ii) one introduces a scalar particle 
with a mass of the order of the electroweak (EW) energy scale which breaks the lepton number
in the scalar sector. A left-handed neutrino mass is then induced by a radiative 
correction from a scalar loop without the right-handed neutrinos.
This model requires some new physics at the EW  scale. 

 Anyway, models of (i) and (ii) reduce to the effective dimension-five operator
 \begin{equation}
   {\kappa_{ij}\o \Lambda}\phi^0 \phi^0 \n_i \n_j   ,
 \end{equation}
 \noindent where $\phi^0$ is the SU(2) doublet Higgs in the SM, which
generates Majorana neutrino masses and mixings.
 The structure of the $\kappa_{ij}/\Lambda$ depend on  details of models \cite{Ma}.
 In the followings, we present typical mass matrix models which lead to the large flavor mixing
 of the  atmospheric(solar) neutrinos.

 {\bf See-saw enhancement:}	
 
 We begin with discussing the see-saw enhancement.
 The see-saw mechanism of neutrino mass generation gives a very natural and
 elegant understanding for the smallness of neutrino masses.
 This mechanism may play another important role, which is to reproduce
  the large flavor mixing.
  In the standpoint of the quark-lepton unification, the Dirac mass
  matrix of neutrinos is similar to the  quark mass matrices.
  Therefore, the neutrino mixings are expected to be typically of the same order of 
  magnitude as the quark mixings.  However, the large flavor mixings of neutrinos could be
  obtained in the see-saw mechanism as a consequence of a certain structure of the
  right-handed Majorana mass matrix \cite{en1,en2}.  
  That is the so called see-saw enhancement of the neutrino
  mixing due to the cooperation between the Dirac and Majorana mass matrices.

  Mass matrix of light Majorana neutrinos $m_\n$ has the following form
  \begin{equation}          
   m_\n \simeq -m_D M_R^{-1} m_D^T \ ,      
  \end{equation}          
 \noindent  
 where $m_D$ is the neutrino Dirac mass matrix and $M_R$ is the Majorana mass matrix
 of the right-handed neutrino components.
 Then, the lepton mixing matrix is \cite{en1}            
   $U_{\rm MNS} = S_\ell^\dagger \cdot S_\n \cdot U_{ss}$,       
  where $S_\ell$, $S_\n$ are transformations which diagonalize the Dirac mass
  matrices of charged leptons and neutrinos, respectively.
  The $U_{ss}$ specifies the effect of the see-saw mechanism, i.e. the effects of
  the right-handed Majorana mass matrix.  It is determined by
   \begin{equation}          
   U_{ss}^T m_{ss}U_{ss} = diag(m_1,m_2,m_3)  \quad {\rm with}\quad 
        m_{ss} = -m_D^{diag} M_R^{-1} m_D^{diag} \ .       
  \end{equation}  
  \noindent
  In the case of two generations,  the mixing matrix $U_{ss}$  is easily investigated
   in terms of one angle $\th_s$.
   This angle could be maximal under the some conditions of parameters
   in the Dirac mass matrix and right handed Majorana mass matrix.
   That is the enhancement due to the see-saw mechanism.
   The rich structure of right-handed  Majorana mass matrix can lead to
    the maximal flavor mixing of neutrinos.
	The detail studies have been given recently in ref. \cite{Jeza,Alt2}.

 {\bf Asymmetric mass matrix:}
 
  The large mixing angle could be derived from the asymmetric mass matrix of charged leptons.
  In the standpoint of the quark-lepton unification, the charged lepton mass matrix
  is connected with  the down quark one.
  The mixing following from the charged lepton mass matrix may be considered to be
   small like  quarks in the hierarchical base.  However, this  expectation
   is not true if the mass matrix is non-Hermitian (asymmetric mass matrix).
   In the SU(5), fermions belong {\bf 10} and {\bf 5*}:
 \begin{equation}    
  {\bf 10}:\  \chi_{ab} =u^c + Q + e^c , \qquad {\bf 5^*}:\  \psi^a  =  d^{c1} + L, 
  \end{equation}          
\noindent where $Q$ and $L$ are SU(2) doublets of  quarks and leptons, respectively. 
 The Yukawa couplings are given by $10_i 10_j 5_H$(up-quarks) and
 $5^*_i 10_j 5^*_H$(down-quarks and charge leptons)(i,j=1,2,3).
 Therefore we get $m_E= m_D^T$ at the GUT scale.
 
 It should be noticed that observed quark mass spectra and the CKM matrix
 only constrains the down quark mass matrix typically as follows:
 \begin{equation}         
   m_{\rm down} \sim  K_D \left (\matrix{    
                \l^4 & \l^3 & \l^4 \cr     
                  x  & \l^2 & \l^2 \cr    
                  y & z  &   1   \cr } \right )    \quad {\rm with} \quad \l=0.22 \ .
      \end{equation}        
\noindent    
 Three parameters   $x,\ y,\ z$ are not determined by observed quark mass spectra 
   and the CKM matrix.  Those are related to the left-handed charged lepton mixing
    due to  $m_E= m_D^T$.
	The left(right)-handed down quark mixings are related to
   the right(left)-handed charged lepton mixings in the SU(5).
	Therefore, there is a source of the large flavor mixing
	in the charged lepton sector if  $z\simeq 1$ or/and  $y\simeq 1$ is derived from some models
	as follows:
	\begin{equation}         
   m_{\ell} m_{\ell}^\dagger = m_{\rm down}^\dagger  m_{\rm down}
    \sim  K_D^2 \left (\matrix{    
                x^2+y^2 & yz+\l^2 x & y+\l^2 x \cr     
                yz+\l^2 x  & z^2 & z \cr    
                y+ \l^2 x & z  &   1   \cr } \right )  \ .
      \end{equation}  
 \noindent
	This mechanism was  used by some authors \cite{SY,Alb,SU5}.

	{\bf Radiative neutrino mass:} 
	
    In the class of models in (ii),
   neutrino masses are induced from the radiative corrections even if the right-handed neutrino
   is absent.
   The typical one is the Zee model, in which charged gauge singlet scalar
   induces the neutrino mass \cite{Zee}.  The diagonal terms of the Zee mass matrix are
   exactly zero due to the symmetry  as follows:
   \begin{equation}         
   m_{\nu} \sim \left (\matrix{0 &  m_{e\mu} & m_{e\tau} \cr     
                  m_{e\mu} & 0&  m_{\mu\tau} \cr    
                  m_{e\tau} & m_{\mu\tau} &   0   \cr } \right )   \ .
      \end{equation}        
   In the case of  $m_{e\mu}\simeq m_{e\tau}\gg m_{\mu\tau}$,
    both solar neutrino problem and atmospheric neutrino deficit can be explained.
	Then, the inverse hierarchy $m_1\simeq m_2 \gg m_3$ and 
   the bi-maximal mixing matrix are obtained \cite{ZeeNew}.
   
   The MSSM with R-parity violation can also give the neutrino masses and mixings
   \cite{KN,Rparity}.
   The MSSM allowes renormalizable B and L violation.  
   The R-parity conservation forbids the B and L violation in the superpotential
    in order to avoid the proton decay.  However 
	the proton decay is avoided in the tree level if either of B or L violating term vanishs.
	The simplest model is the bi-linar R-parity violating model with $\epsilon_i H_u L_i$ 
	for the lepton-Higgs coupling \cite{KN}.
	This model provides the large mixing which is consistent with 
	atmospheric and solar neutrinos.

 \section{Search for Flavor Symmetry}
 
  Masses and mixings of the quark-lepton may suggest the some flavor symmetry.
  The simple flavor symmetry is U(1), which was discussed intensively by
  Ramond et al.\cite{Ramond}.
  In their model, they assumed
  (1) Fermions carry U(1) charge, (2) U(1) is spontaneously broken by $<\theta>$, in which
  $\theta$ is the EW singlet with U(1) charge -1, and (3) Yukawa couplings appear
   as effective operators a la Froggatt-Nielsen mechanism \cite{FN},
  \begin{equation}          
       h^D_{ij} Q_i \bar d_j H_d \left ({\theta\o \Lambda}\right )^{m_{ij}} + 
	   h^U_{ij} Q_i \bar u_j H_u \left ({\theta\o \Lambda}\right )^{n_{ij}} +...\ ,     
  \end{equation} 
  \noindent
   where $<\theta>/ \Lambda=\lambda\simeq 0.22$.
   The powers ${m_{ij}}$ and  ${n_{ij}}$ are determined from the U(1) charges
   of  fermions in order that the  effective operators are  U(1) invariants as,
    \begin{equation}          
       m_{ij}=Y_{Q_i}+Y_{\bar d_j}+Y_{H_d} \ , \qquad  n_{ij}=Y_{Q_i}+Y_{\bar u_j}+Y_{H_u}\ ,     
  \end{equation} 
  \noindent
  where $Y$ denotes the U(1) charge.
   The U(1) charges of the fermions are fixed by the experimental data of the
   fermion masses and mixings. 
   Their naive U(1) symmetric mass matrices could be modified by
   taking account of new fields or new symmetries.


Another approach is based on the non-Abelian flavor symmetry  $S_3$.
  The $S_{3L}\times S_{3R}$ symmetric mass matrix is so called 
 the democratic mass matrix \cite{Demo},
 \begin{equation}
M_q= {c_q}  \left( \matrix{1 & 1 & 1 \cr
                            1 & 1 & 1 \cr
                            1 & 1 & 1 \cr  } \right) \ ,
\end{equation}
\noindent
 which  needs the large rotation in order to move to
  the diagonal base as $A^T M_q  A$, where
  \begin{equation}
A= \left ( \matrix{1/\sqrt 2 & 1/\sqrt 6 & 1/\sqrt 3 \cr
                   -1/\sqrt 2 & 1/\sqrt 6 & 1/\sqrt 3 \cr
                           0 & -2/\sqrt 6 & 1/\sqrt 3 \cr   } \right ) \ .
 \label{Orth}
 \end{equation}
  In the CKM mixing matrix, this large rotation matrix $A$ is completely canceled each other 
  between down quarks and up quarks. This democratic mass matrix is not a realistic one
  because two quarks are massless.
  There are many works in which realistic quark mass matrices 
   are discussed including  symmetry breaking terms in the quark sector \cite{Koide}.
  However, the situation of the lepton sector is very different from
  the quark sector since the effective neutrino mass matrix $m_{LL}^\n$ could be 
  far from the democratic one and the charged lepton one is still the democratic one. 
  
 The neutrino mass matrix is different from the democratic one if they are Majorana particles.
The $S_{3L}$ symmetric mass term is given as follows:
\begin{equation}
M_\n= {c_\n}  \left( \matrix{1 & 0 & 0 \cr
                            0 & 1 & 0 \cr
                            0 & 0 & 1 \cr  } \right)
	+ {c_\n} r \left( \matrix{0 & 1 & 1 \cr
                            1 & 0 & 1 \cr
                            1 & 1 & 0 \cr  } \right) \ , 
  \end{equation}
\noindent
where $r$ is an arbitrary parameter.
The  eigenvalues of this matrix are given as 
$c_\n(1+2r, \ 1-r, \ 1-r)$, which means that there are at least two degenerate masses
 in the $S_{3L}$ symmetric Majorana mass matrix \cite{FTY,KK,Lisbon}.
 
 In order to explain both solar and atmospheric neutrinos, three neutrinos
 should be almost degenerate in this model.
 If three degenerate light neutrinos are required, the parameter $r$ should be
  taken as $r=0$ or $r=-2$.  The first case was discussed in ref.\cite{FTY}
   and the second case was discussed in ref.\cite{KK}.
   The difference of $r$ leads to the difference in the $CP$ property of  neutrinos.
   
In order to reproduce the atmospheric neutrino deficit by the large neutrino
oscillation, the symmetry breaking terms are required.
Since  results are almost same, we show the numerical analyses
  in ref.\cite{FTY}.

 Let us start with  discussing the following  charged lepton mass matrix:
 \begin{equation}
 M_\ell= {c_\ell \over 3}
             \left ( \matrix{1 & 1 & 1 \cr
                            1 & 1 & 1 \cr
                            1 & 1 & 1 \cr
                                         } \right )
+ m_{\rm break} \ .
  \end{equation}
\noindent
 The first term is a unique representation of
the $S_{3L}\times S_{3R}$ symmetric matrix and the second one $m_{\rm break}$
is a symmetry breaking one.
The unitary matrix that diagonalises the charged lepton mass matrix is  $U_\ell=AB_\ell$,
 where the matrix $A$ is defined in eq.(\ref{Orth}) and
$B_\ell$ depends on the symmetry breaking term $m_{\rm break}$.

Let us turn to the neutrino mass matrix, in which $r=0$ is taken:
\begin{equation}
M_\n= {c_\n}
             \left ( \matrix{1 & 0 & 0 \cr
                            0 & 1 & 0 \cr
                            0 & 0 & 1 \cr  } \right )
+\left (\matrix{0 & \epsilon_\nu & 0 \cr
                 \epsilon_\nu  & 0 & 0 \cr
                  0 & 0 & \delta_\nu \cr} \right ) \ ,
  \label{neumass}
  \end{equation}
\noindent
where the symmetry breaking is given by  a small term with two adjustable parameters. 
It is remarked that $S_{3L}$ is broken by $\delta_\nu$ but  $S_{2L}$ is still preserved
in eq.(\ref{neumass}).
The mass eigenvalues 
are $c_\nu\pm\epsilon_\nu$, and 
$c_\nu+\delta_\nu$, and the matrix
that diagonalises $M_\nu$  ($U_\nu^TM_\nu U_\nu=$diagonal) is

\begin{equation}
U_\nu= \left ( \matrix{1/\sqrt 2 & 1/\sqrt 2 & 0 \cr
                   -1/\sqrt 2 & 1/\sqrt 2 & 0 \cr
                           0 & 0 & 1 \cr
                                         } \right ) \ .
\end{equation}
 That is, our $M_\nu$ represents three degenerate neutrinos, with the degeneracy
lifted by a small parameters. 

The lepton mixing angle 
as defined by $U_{\rm MNS}=(U_\ell)^\dagger U_\nu=(AB_\ell)^\dagger U_\nu$ is thus given
by

\begin{equation}
U_{\rm MNS}\simeq
\left ( \matrix{1 & \frac{1}{\sqrt{3}} B_{\ell 21}  & -\frac{2}{\sqrt{6}}B_{\ell 21}  \cr
        B_{\ell 12}  & \frac{1}{\sqrt{3}} &  -\frac{2}{\sqrt{6}} \cr
                           0 & \frac{2}{\sqrt{6}} & \frac{1}{\sqrt{3}} \cr } \right ) \ , 
\end{equation}
where $B_{\ell 21}$ and  $B_{\ell 12}$ are correction terms in the charged lepton sector,
typically, $B_{\ell 21}\sim \sqrt{m_e/m_\mu}$.
We have predictions
\begin{equation}
  \sin^22\theta_{\rm atm}\simeq \frac{8}{9} \ , \qquad U_{e2}\simeq -\frac{1}{\sqrt{2}} U_{e3} \ ,
\end{equation}
\noindent where $U_{\a i}$ denotes the MNS mixing.
If $B_{\ell 21}\simeq \sqrt{m_e/m_\mu}$, we get
  $U_{e2}\simeq 0.04$ and  $U_{e3}\simeq 0.057$, which leads to 
$\sin^22\theta_\odot\simeq 6.5\times 10^{-3}$. This prediction also agrees with the neutrino 
mixing corresponding to the
small mixing angle MSW solution $(4 -13) \times 10^{-3}$ 
 for the solar neutrino problem \cite{BKS}.
  In the future, this prediction will be tested in the
following  long baseline experiments $\n_\n \Ar \n_e$  and $\n_e \Ar \n_\tau$.
 

Let us briefly discuss the
consequence of the other symmetry breaking of neutrino masses.
If we adopt the symmetry breaking term alternative to eq.(\ref{neumass}), 
\begin{equation}
  \left ( \matrix{\rho_\nu & 0 & 0 \cr
                 0  & \epsilon_\nu & 0 \cr
                  0 & 0 & \delta_\nu \cr} \right ) \ ,
\end{equation}
in which $S_{3L}$ is completely broken,
we obtain the lepton mixing matrix to be
\begin{equation}
U_{\rm MNS}\simeq A^T= \left ( \matrix{1/\sqrt 2 & -1/\sqrt 2 & 0 \cr
                   1/\sqrt 6 & 1/\sqrt 6 &  -2/\sqrt 6 \cr
                           1/\sqrt 3 &  1/\sqrt 3 &  1/\sqrt 3 \cr
                                         } \right ) \ .
 \end{equation}
\noindent
 This is identical to the matrix presented by Fritzsch and Xing \cite{FX}.
  For this case one gets
\begin{equation}
\sin^22\theta_\odot\simeq 1,\qquad \qquad \sin^22\theta_{\rm atm}\simeq 8/9 \ .
\end{equation}
\noindent
This case can accommodate the "just-so" solution 
 for the solar neutrino problem due to neutrino oscillation in vacuum 
  and may be also consistent with the large mixing angle MSW solution
  including correction terms.
This matrix has been investigated in detail \cite{Tanimoto} focusing on recent data
 at Super-Kamiokande.

In the model, the symmetry breaking terms are not unique, and moreover,
the neutrino mass degeneracy is put by hand, $r=0$.
In order to avoid these ambiguity, we should go to higher symmetry of flavors.

\section{$\bf O(3)$ Flavor Symmetry and Phenomenology} 

  We assume that  neutrinos are almost degenerated. 
  Since the quark-lepton masses are hierarchical, one may raise a question.
  How can one gets the  consistent picuture in these mass generation?
  The $O(3)$ flavor symmetry \cite{O31,O32}
 has a unique prediction, that is almost degenerate neutrino masses.
 Masses of quarks and charged leptons vanish
   in the $O(3)$ symmetric limit.
   Therefore, mass matrices of quarks and leptons are determined by details of
   breaking pattern of the flavor symmetry.  
   Although there are some symmetry breaking mechanism \cite{O31,O32}
   we discuss a possible flavor $O(3)$ breaking mechanism \cite{TWY} that leads to
   "successful" phenomenological mass matrices with $S_3$ symmetry in the previous section.
   
   We consider the supersymmetric standard model and impose
  $O(3)_L\times O(3)_R$ flavor symmetry.  Three lepton doublets $\ell_i(i=1-3)$
transform as an $O(3)_L$ triplet and three charged leptons $\bar e_i(i=1-3)$
	as an $O(3)_R$ triplet, while Higgs doublets $H$ and $\bar H$ are 
 $O(3)_L\times O(3)_R$ singlets.  We will discuss the quark sector later.
	
	 We introduce, to break the flavor symmetry, pair of fields
	 $\Sigma^{(i)}_{L} (i=1,2)$ and  $\Sigma^{(i)}_{R} (i=1,2)$ 
	 which transform as symmetric traceless tensor {\bf 5}'s of  $O(3)_L$
	 and  $O(3)_R$, respectively.
	 We assume that the $\Sigma^{(i)}_{L}({\bf 5},{\bf 1})$ and 
	  $\Sigma^{(i)}_{R}({\bf 1},{\bf 5})$ take values
	\begin{equation}
\Sigma^{(1)}_{L,R} =
\left( \matrix{1 & 0 & 0 \cr
        0 & 1 & 0 \cr  0 & 0 & -2  \cr  } \right )  w^{(1)}_{L,R} \ ,
	\label{S1}
\end{equation}		
\noindent
and 					 
\begin{equation}
\Sigma^{(2)}_{L,R} =
\left( \matrix{1 & 0 & 0 \cr
        0 & -1 & 0 \cr  0 & 0 & 0  \cr  } \right )  w^{(2)}_{L,R}\ .
\label{S2}
\end{equation}		
We consider that these are explicit breakings  of
	$O(3)_L\times O(3)_R$  rather than vacuum-expectation values of  
	$\Sigma^{(i)}_{L,R}$(spontaneous breaking),
	otherwise we have unwanted massless Nambu-Goldstone multiplets.
	In the following discussion we use dimentionless breaking parameters 
	$\sigma^{(i)}_L$ and $\sigma^{(i)}_R$, which are defined as
\begin{equation}
 \sigma^{(1)}_{L,R}\equiv \frac{\Sigma^{(1)}_{L,R}}{M_f} =
  \left( \matrix{1 & 0 & 0 \cr
        0 & 1 & 0 \cr  0 & 0 & -2  \cr  } \right )  \delta_{L,R} \ ,
\end{equation}	
\noindent  and 		
 \begin{equation}
  \sigma^{(2)}_{L,R}\equiv \frac{\Sigma^{(2)}_{L,R}}{M_f} =
  \left( \matrix{1 & 0 & 0 \cr
        0 & -1 & 0 \cr  0 & 0 & 0 \cr  } \right )  \e_{L,R} \ .
 \end{equation}	
 \noindent Here, $M_f$ is the large flavor mass scale,
 $\delta_{L,R}=w_{L,R}^{(1)}/M_f$ and $\e_{L,R}=w_{L,R}^{(2)}/M_f$.
 We assume $\delta_{L,R}, \e_{L,R} \leq 1$.
  
   The neutrinos acquire small Majorana masses from a superpotential,
 \begin{equation}
  W =\frac{H^2}{M}\ell ( {\bf 1}+\a_{(i)}\sigma^{(i)}_L ) \ell  \ ,
\end{equation}	
\noindent
which yields a neutrino mass matrix as
\begin{eqnarray}
  \widehat m_\n =
  \frac{<H>^2}{M}\Biggl \{  \left (\matrix{1 & 0 & 0 \cr
        0 & 1 & 0 \cr  0 & 0 & 1 \cr  }  \right )  
		 &+& \a_{(1)} \left (\matrix{1 & 0 & 0 \cr
                0 & 1 & 0 \cr  0 & 0 & -2 \cr  }   \right ) \delta_L   \nonumber \\
		&+& \a_{(2)} \left (\matrix{1 & 0 & 0 \cr
        0 & -1 & 0 \cr  0 & 0 & 0 \cr  } \right ) \e_L \Biggr \}  .
	\label{nuemass0}
 \end{eqnarray}
 \noindent Here, $\a_{(i)}$ are ${\cal O}(1)$ parameters and the mass  $M$ denotes
  a cut-off scale of the present model which may be different from the flavor scale $M_f$.
  We take $M\simeq 10^{14-15} \G$ to obtain 
  $m_{\n_i}\simeq 0.1- 1 \eV$ indicated from the atmospheric neutrino oscillation
  \cite{SKam} for degenerate neutrinos.
  
  The above breaking is, however, incomplete, since the charged leptons
  remain massless. We introduce an $O(3)_L$-triplet and an
   $O(3)_R$-triplet fields
   $\phi_L({\bf 3},{\bf 1})$ and $\phi_R({\bf 1},{\bf 3})$ to produce masses of
   the charged leptons.  The vacuum expectation values of 
   $\phi_L$ and $\phi_R$ are determined by the following superpotential;
 \begin{eqnarray}
  W &=& Z_L (\phi^2_L - 3v_L^2) + Z_R (\phi^2_R - 3v_R^2) \nonumber \\
     && +  X_L ( a_{(i)} \phi_L \sigma_L^{(i)} \phi_L )  
	  +  X_R ( a'_{(i)}\phi_R \sigma_R^{(i)} \phi_R )  \nonumber \\
	  &&+  Y_L ( b_{(i)} \phi_L \sigma_L^{(i)}   \phi_L)  
	  +  Y_R ( b'_{(i)} \phi_R \sigma_R^{(i)} \phi_R)  \ .
  \label{Super}
 \end{eqnarray}	
\noindent
Here, the fields $Z_{L,R}$, $X_{L,R}$ and $Y_{L,R}$ are all singlets 
of $O(3)_L\times O(3)_R$.

We obtain vacuum-expectation values from the superpotential eq.(\ref{Super})
by solving $|F_X|=0$, $|F_Y|=0$ and $|F_Z|=0$:
 \begin{equation}
  <\phi_L> \equiv \left ( \matrix{1\cr 1 \cr 1 \cr} \right ) v_L \ , \qquad
  <\phi_R> \equiv \left ( \matrix{1\cr 1 \cr 1 \cr} \right ) v_R \  .
  \label{vac}
 \end{equation}
 Notice that only with the first two terms in eq.(\ref{Super}) we have 
 $O(3)_L\times O(3)_R$ global symmetry and hence unwanted Nambu-Goldstone multiplets appear
  in broken vacua.  The couplings to  the explicit breakings 
  $\sigma_{L,R}^{(i)}$ are necessary to eliminate the Nambu-Goldstone multiplets
  in the low energy spectrum, which determine  vacuum-expectation values 
  of $\phi_L$ and $\phi_R$ as in eq.(\ref{vac}). 
  
   With the non-vanishing $<\phi_L>$ and  $<\phi_R>$
   in eq.(\ref{vac}), the Dirac masses of charged leptons
   arise from a superpotential,
   \begin{equation}
      W = \frac{\k_E}{M_f^2} (\bar e\phi_R) (\phi_L \ell)\bar H  .
   \end{equation}
   \noindent
   This produces so-called "democratic" mass matrix of the charged leptons,
   \begin{equation}
  \widehat m_E = \k_E \left ( \frac{v_L v_R}{M_f^2}\right )
    \left (\matrix{1 & 1 & 1 \cr
        1 & 1 & 1 \cr  1 & 1 & 1 \cr }  \right ) <\bar H> .
		\label{lmass0}
  \end{equation}
	Diagonalization of this mass matrix yields large lepton mixings \cite{FTY,FX}
	 and one non-vanishing eigenvalue, $m_\tau$.
	The masses of e and $\mu$ are derived from distortion of the 
	"democratic" form of mass matrix in eq.(\ref{lmass0}), which is
	given by a superpotential containing the explicit 
	$O(3)_L\times O(3)_R$ breaking parameters  $\sigma_{L,R}^{(i)}$,
 \begin{eqnarray}
  \delta W = \frac{\k_E}{M_f^2} \Biggl \{  A_i^\ell (\bar e \sigma^{(i)}_R\phi_R) (\phi_L \ell)
   &+& B_i^\ell (\bar e \phi_R) (\phi_L \sigma^{(i)}_L \ell) \nonumber \\
   &+& C_{ij}^\ell (\bar e \sigma^{(i)}_R \phi_R) (\phi_L \sigma^{(j)}_L \ell) \Biggr \} \bar H .
   \label{dW}
   \end{eqnarray}
   \noindent
   Then, the charged lepton mass matrix is given in the hierarchical base by
\begin{equation}
  A^T \widehat m_E  A = 
   \frac{\k_E v_L v_R}{M_f^2}  <\bar H>
  \left ( \matrix{2 C_{22}^\ell\e_L \e_R & 2\sqrt{3} C_{21}^\ell\e_R\d_L  
                          & \sqrt{6} A_2^\ell \e_R \cr
						  &        &        \cr
	2\sqrt{3}  C_{12}^\ell \e_L \d_R & 6 C_{11}^\ell \d_L \d_R  & 3\sqrt{2} A_1^\ell \d_R  \cr 
	                      &            &             \cr
	 \sqrt{6} B_2^\ell \e_L  & 3\sqrt{2} B_1^\ell\d_L  & 3  \cr  
	 } \right )_{RL}  
\label{lmassh}
\end{equation}
 \noindent where the matrix $A$ is defined in eq.(\ref{Orth}).
The mass eigenvalues of this lepton mass matrix are
 \begin{equation}
  m_\tau \simeq 3\k_E \frac{v_L}{M_f}\frac{v_R}{M_f}<\bar H> , \qquad
  \frac{m_\m}{m_\tau} \simeq {\cal O}(\d_L \d_R), \qquad 
  \frac{m_e}{m_\tau} \simeq {\cal O}(\e_L \e_R),
 \end{equation}
\noindent where we assume that all coupling parameters 
$A_i^\ell$, $B_i^\ell$ and $C_{ij}^\ell(i,j=1,2)$ are of ${\cal O}(1)$.

We now turn to the quark sector, in which three doublet quarks
 $q_i$ transform as an $O(3)_L$ triplet while three down quarks $\bar d_i$
 and the three up quarks $\bar u_i$ as $O(3)_R$ triplets. 
Quark mass matrices are same ones in  eq.(\ref{lmassh}) apart from ${\cal O}(1)$ coefficients 
 $A_i^\ell$, $B_i^\ell$ and $C_{ij}^\ell$.
 The CKM mixing angles are given by
  \begin{equation}
  \left |V_{us} \right | \simeq \frac{\e_L}{\d_L} \ ,\qquad
  \left |V_{cb} \right | \simeq \delta_L \ ,\qquad
  \left |V_{ub} \right | \simeq \e_L \ .
  \end{equation}
 Putting the experimental quark mass ratios and CKM matrix elements:
  \begin{equation}
  \frac{m_d}{m_b}\simeq \l^4, \quad \frac{m_s}{m_b}\simeq \l^2, 
   \quad \left |V_{us} \right | \simeq \l,  \quad 
                                  \left |V_{cb} \right | \simeq \l^2 ,
 \end{equation}
 we obtain the order of parameters as follows:
 \begin{equation}
  \d_L \simeq \l^2, \qquad \d_R\simeq 1 \ , \qquad
  \e_L \simeq \l^3, \qquad \e_R \simeq \l \  ,
  \label{mag}
 \end{equation}
 \noindent  with $\l \simeq 0.2$.
 Then, we predict  $\left |V_{ub} \right |\simeq \e_L \simeq \l^3$,
  which is consistent with the experimental value \cite{PDG}. 
  Thus our model is successful to explain  both lepton and quark mass matrices.

 Let us discussing neutrino masses and the  mixings.	
 Following from the analysis on the quark mass matrices we take 
 $\d_L\simeq 0.1$ and $\e_L\simeq 10^{-3}-10^{-2}$.
  We should remark that there is an additional contribution to the neutrino mass matrix 
  in eq.(\ref{nuemass0}) as
   \begin{equation}
  \delta W = \frac{H^2}{M}\ell\left ( \b \frac{\phi_L}{M_f} \frac{\phi_L}{M_f}\right )\ell .
   \end{equation}
   \noindent
The neutrino mass matrix is now given by 
\begin{eqnarray}
  \widehat m_\n &=&
  \frac{<H>^2}{M}\Biggl \{ \left (\matrix{1 & 0 & 0 \cr
        0 & 1 & 0 \cr  0 & 0 & 1 \cr  }  \right ) + 
		 \a_{(1)} \left (\matrix{1 & 0 & 0 \cr
        0 & 1 & 0 \cr  0 & 0 & -2 \cr  }   \right ) \delta_L \nonumber \\
		 &+& \a_{(2)} \left (\matrix{1 & 0 & 0 \cr
        0 & -1 & 0 \cr  0 & 0 & 0 \cr  } \right ) \e_L 
		 +\b \left (\frac{v_L}{M_f}\right )^2 
		\left (\matrix{1 & 1 & 1 \cr  1 & 1 & 1 \cr  1 &  1 &  1 \cr  }\right )
		 \Biggr \} .
	\label{nuemass}
 \end{eqnarray}
 \noindent
 The large MNS mixing angle between  $\n_\m$ and $\n_\tau$ 
  is obtained if 
\begin{equation}
 \b \left (  \frac{v_L}{M_f}\right )^2 \ll \a_{(1)} \d_L  \ .
 \label{cond}
\end{equation}
 We also see  large  neutrino mixings between $\n_e$ and $\n_{\m, \tau}$
 for $\b (v_L/M_f)^2\leq \a_{(2)}\e_L$.
 By using 
 $\Delta m^2_{23}(\equiv m_{\n_3}^2-m_{\n_2}^2) \simeq 10^{-3} \eV^2$  for the $\n_\mu-\n_\tau$
 oscillation \cite{SKam} (which corresponds to $m_{\n_i}={\cal O}(0.1) \eV$), 
 $\d_L\simeq 0.1$ and  $\e_L\simeq 10^{-3}-10^{-2}$, we obtain
 \begin{equation}
 \Delta m^2_{12} \simeq \frac{\e_L}{\d_L}\Delta m^2_{23} \simeq 10^{-5}-10^{-4} \eV^2 ,
 \end{equation}
 \noindent for the $\n_e-\n_{\mu,\tau}$ oscillation.
 This  is consistent with the large angle MSW solution \cite{MSW}
 to the solar neutrino problem.
The current analyses \cite{BKS2} of Super-Kamiokande experiments
give $\Delta m^2_{12} \simeq 2\times 10^{-5}-2\times 10^{-4} \eV^2$ and
 $\sin^2 2\theta_{12}=0.60-0.97$ at the $99\%$ confidence level, for the
large MSW  solution.
 It is remarked that we obtained the numerical prediction
  $\sin^2 2\theta_{12}=0.60-0.97$  under the condition $\b (v_L/M_f)^2\leq \a_{(2)}\e_L$.

 
 We  considered a model where $\ell_i$ and $q_i$ belong to
 triplets of one $O(3)$ and $\bar e_i$, $\bar d_i$ and  $\bar u_i$ belong to
 triplets of the other  $O(3)$.  We note here that  there is another
 interesting assignment that $\ell_i$ and $\bar d_i$ are triplets of the 
 $O(3)$ while $\bar e_i$, $q_i$ and  $\bar u_i$ transform as triplets of the other $O(3)$
 by imposing a discrete symmetry such as $Z_6$ \cite{TWY}. 

\section{Summary}
  We have presented some typical mechanism to leads
  models for the lepton  mass matrix, which give the near-maximal flavor mixing.
    Especially, 
     details of the models with the  $S_3$ or $O(3)$ flavor symmetry are presented.
	 Since these models predict almost degenerated neutrino masses, double-$\b$ decay
	  experiments will test the model in the future \cite{Double}.
	Our $O(3)_L\times O(3)_R$ model predicts the large mixing angle MSW solution,
	we wait for results in KamLAND experiment \cite{KAMLAND}.
	More theoretical works as to the flavor symmetry as well as experimental data
	 are expected.

\vskip 0.5 cm
 \section*{Acknowledgements}

  I thank T. Watari and  T. Yanagida for collaboration on the lepton mass matrix model
   with the  $O(3)$ symmetry. 
  This research is  supported by the Grant-in-Aid for Science Research,
 Ministry of Education, Science and Culture, Japan(No.10640274).

\end{document}